\documentclass[%
reprint,
 amsmath,amssymb,
 aps,
 prl,
]{revtex4-2}
\usepackage{graphicx}
\usepackage[dvipsnames]{xcolor}
\usepackage{dcolumn}
\usepackage{bm}
\usepackage{mathtools}

\usepackage{soul}       

\graphicspath{{./}}

\begin{document}

\preprint{APS/123-QED}

\title{
Insensitive edge solitons in a non-Hermitian topological lattice
}

\author{Bertin Many Manda}
 \email{bertin.many\_manda@univ-lemans.fr}
\author{Vassos Achilleos}%
 \email{achilleos.vassos@univ-lemans.fr}
\affiliation{%
  Laboratoire d’Acoustique de l’Universit\'e du Mans (LAUM), UMR 6613, Institut d'Acoustique - Graduate School (IA-GS), CNRS, Le Mans Universit\'e, Av. Olivier Messiaen, 72085 Le Mans, France 
}%

\begin{abstract}

In this work, we demonstrate that the synergetic interplay of topology, nonreciprocity and nonlinearity is capable of unprecedented effects.
We focus on a nonreciprocal variant of the Su-Shrieffer-Heeger chain with local Kerr nonlinearity. 
We find a continuous family of non-reciprocal edge solitons (NES) emerging from the topological edge mode, with near-zero energy, in great contrast from their reciprocal counterparts. 
Analytical results show  that this energy decays exponentially towards zero when increasing the lattice size.
Consequently, despite the absence of chiral and sublattice symmetries within the system, we obtain zero-energy NES, which are insensitive to growing Kerr nonlinearity.
Even more surprising, these zero-energy NES also persist in the strong nonlinear limit. 
Our work may enable new avenues for the control of nonlinear topological waves
 without requiring the addition of complex chiral- or sublattice-preserving nonlinearities.
\end{abstract}

\maketitle



During the last two decades, great progress has been achieved in understanding topological  systems~\cite{HK2010,MHSP2014,MXC2019,OPAGHLRSSZC2019,MHKIEMKKTBS2020,NYFRADNHH2020,SB2021,ZZCLC2023,SGXQGOEMF2024}.
The promise is the developments of technological devices exploiting localized edge waves immune to fluctuations and defects. 
The manifestations of these topological edge modes were observed in experiments in photonic~\cite{OPAGHLRSSZC2019,SLCK2020,JJMR2024,SR2024}, atomic~\cite{MADMHG2018,LWPLBSS2019}, electronic~\cite{DJVR2021,GJPLDF2024} and phononic~\cite{XSHMY2020,ZDX2022,BMMM2024,LMLYTD2024} devices. 
Another domain which has been extensively debated recently are linear waves in non-Hermitian (NH) media~\cite{SISL2020}.
These media offer promising solutions in implementing  unidirectional, broadband waveguides and amplifiers~\cite{BLLC2019}, known to host a plethora of new phenomena like non-Hermitian skin effect (NHSE)~\cite{MBF2018,OKSS2020,LXDLLGYY2022}, complex-frequency and exceptional points~\cite{MAL2024}.
In most cases, the NHSE originates from the asymmetry between the couplings which constraints most of the system 
normal modes (eigenmodes) to be localized on a single interface of the system.
This phenomenon has been experimentally demonstrated in photonics~\cite{WKHHSGTS2020,LLXWYX2022}, electronics~\cite{LSMZYWJJZ2021}, acoustics~\cite{ZYGGCYCXLJYSCZ2021,MAPPA2023} and mechanics~\cite{BLC2019,GBVC2020,WWM2022}.
Owing to this, it was shown that NHSE and topological modes can interact, leading to unprecedented level of manipulation of the latter, e.g., in phononic~\cite{GBVC2020,GWC2020,ZTLG2021,WWM2022} and electric~\cite{LCHWLHDL2024} systems.
 
Interestingly, the inclusion of nonlinear features in NH topological insulators has recently seen growing interest.
Exciting developments like nonlinear NHSE~\cite{Y2021,E2022,KMM2023,JCZL2023,MCKA2024,WWLQZLLL2024}, nonlinearly induced NH phase transitions~\cite{DAMYZZLYTL2024} and nonlinear wave acceleration~\cite{VGGSMC2023} have already been demonstrated.
Furthermore, experiments featuring nonreciprocity revealed the realization of lossless unidirectional waveguides~\cite{LW2022,JS2023,LWWLMJ2023,VGGSMC2023}.
 Following these efforts, we raise the question of what is the fate of nonlinear topological solitons (or breathers)~\cite{ACM2014,HKA2016,LC2016,LRPS2016,PVLR2018,CT2019,CXYKT2021,MS2021,JD2022} in the presence of nonreciprocity?
 Assuming that nonlinearity is perturbative, we can prematurely collect some elements of answers. Indeed, it was demonstrated~\cite{BB2020,YZZWPFSZ2023} that NH topological systems are exponentially sensitive to perturbations at the boundaries. 


\begin{figure}[t!]
    \centering    \includegraphics[width=\columnwidth]{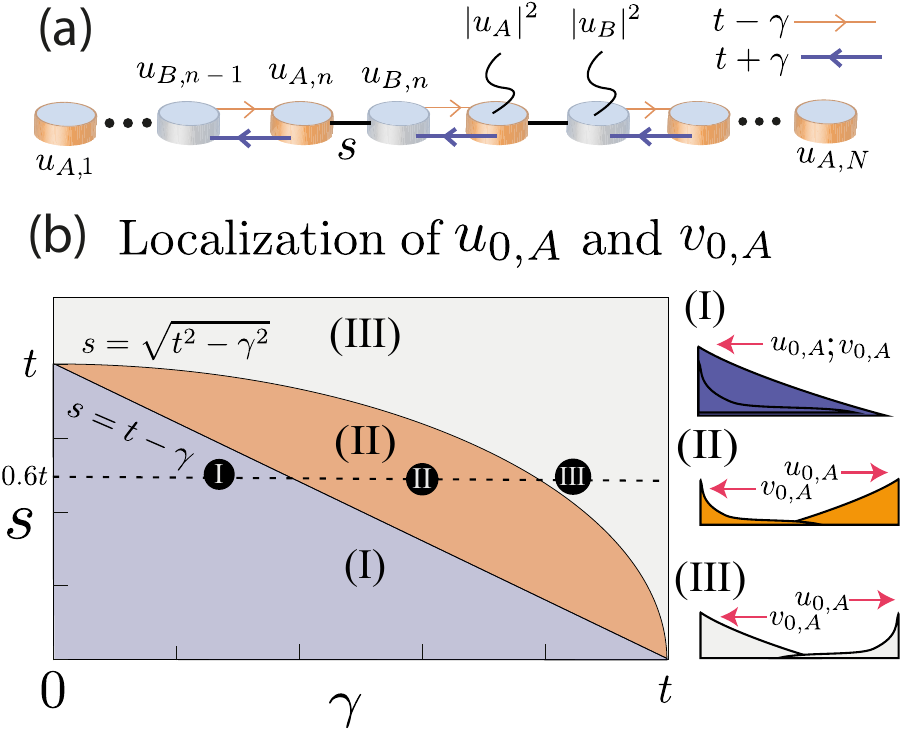}
    \caption{ (a): The nonreciprocal SSH model considered in this work with Kerr nonlinearity. 
      (b)
      Phase diagram $(\gamma, s)$ summarizing the localization characteristics of the TM right and left eigenvectors. 
      Representative cases are shown for $s = 0.6t$ with (I) $\gamma = 0.3t$, (II) $\gamma = 0.6t$, and (III) $\gamma = 0.825t$ (see black circles).
      Right: The localization behavior of the right and left eigenvectors of the TM in regions (I), (II), and (III). 
    }
    \label{fig:fig_linear_01}
\end{figure}
Here we demonstrate an unanticipated insensitivity of the energy of nonlinear edge modes, which we hereby call nonreciprocal edge solitons  (NES) as their intensity increases. 
We consider a Su-Shrieffer-Heeger (SSH) chain~\cite{SSH1979} with nonreciprocal couplings 
featuring Kerr nonlinearity.
As we will explain below, this rather generic model exhibits a region in its parameter space whereby increasing the intensity, $I$, the energy, $E$, of the NES remains practically constant and 
exponentially decays toward $E=0$,
%
with growing lattice size $N$,  
\begin{equation}
    \mathcal{S}_E = \frac{\partial E}{\partial I} \sim \exp \left(-\alpha_E N\right).
\end{equation}





Our results are based on finite lattices whose   governing equations read~\cite{CZLC2022,SM0001}
\begin{equation}
    \begin{split}
            Eu_{A,n} &= \left(t+\gamma\right) u_{B, n-1} + s u_{B, n} + \sigma \left\lvert u_{A,n}\right\rvert^2 u_{A,n} \\
            Eu_{B,n} &= su_{A, n} + \left(t-\gamma\right) u_{A, n+1} + \sigma \left\lvert u_{B,n}\right\rvert^2 u_{B,n}
    \end{split}  ,
    \label{eq:static_equations_of_motion}
\end{equation}
with $E$ being the energy, $\sigma=\pm 1$ the nonlinear coefficient, $s$ and $t\pm\gamma$ the positive nearest-neighbour couplings, and $\gamma>0$ quantifying the strength of nonreciprocity.  
\begin{figure}[t!]
    \centering   \includegraphics[width=\columnwidth]{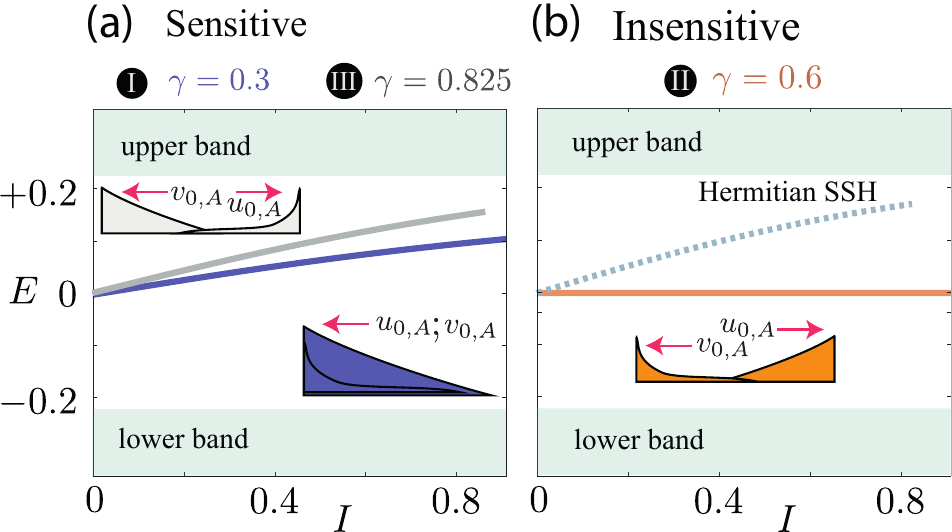}
    \caption{
    The numerically obtained $E$ of the NES as a function of $I$ for representative cases at $s=0.6$ is shown in (a) for regions (I) and (III) with $\gamma = 0.3$ [blue curve] and $\gamma = 0.825$ [gray curve]. (b) Same as in (a) but for region (II) with $\gamma = 0.6$ [orange curve], see also Fig.~\ref{fig:fig_linear_01}(b).
    The dashed line in (b) represents the corresponding result for the Hermitian SSH model with $\gamma = 0$, for comparison.
    The insets depict the localization characteristics of the right and left eigenvectors of the TM.
    }
    \label{fig:sensitivity_01}
\end{figure}
Here $u_{a,n}$ is the amplitude at cell with index $n$ and sublattice $a=\{A,B\}$, Fig.~\ref{fig:fig_linear_01}(a).
In the linear limit, $\lvert u_{a,n} \rvert^2 \rightarrow 0$,
Eq~\eqref{eq:static_equations_of_motion}, 
 reduces to a linear eigenvalue problem $H \vec{u} = E \vec{u}$, with the $H$ being a non-Hermitian matrix possessing chiral and sublattice symmetries (CS and SLS)~\cite{L2016,YW2018,KEBB2018,KSUS2019,CZLC2022,SM0001}. 
Hereafter, we consider a chain of $N$ cells with
$2N-1$ sites and open boundary condition (OBC), Fig.~\ref{fig:fig_linear_01}(a).

The SLS/CS ensures all bulk eigenenergies to come in pairs 
$E_l=\pm \sqrt{s^2 + \tilde{t}^2 + 2s\tilde{t}\cos \left(\frac{l\pi}{2N} \right)}$ with $\lvert l\rvert =1, 2,\ldots, N-1$, labeling the eigenmodes energies and $\tilde{t}=\sqrt{t^2-\gamma^2}$.
Regarding their associated $2N-2$ eigenvectors, non-Hermiticity implies the $l$-th right ($\vec{u}_l$) and left  ($\vec{v}_l$) eigenvectors are in general different~\cite{NOTE0001}.
These eigenvectors exhibit a localised (\emph{skinny}) profile where each site amplitude satisfy 
\begin{equation}
    u_{l,a, n} \sim d_R^{n-1} \mbox{ and } v_{l,a, n} \sim d_L^{n-1},
\end{equation}
with $d_R =d_L^{-1}= \sqrt{(t+\gamma)/(t-\gamma)}$ (see supplementary~\cite{SM0001}).
Thus the right and left eigenvectors of bulk modes have equivalent localization, but are respectively located on the left and right boundaries of the system. 
This localization characteristics, $d_Rd_L=1$, is independent of the parameter coordinates in Fig.~\ref{fig:fig_linear_01}(b).
In addition, the extent of localization solely depends on the non-reciprocal strength, $\gamma$, similar to the Hatano-Nelson model. 
These bulk modes are commonly referred to as skin modes.

In addition to the bulk modes, the  choice of $2N-1$ sites and SLS/CS guarantee the presence of a zero-energy mode with index $l=0$, i.e. $E_{0} = 0$, pinned as topological mode (TM). 
Its left and right eigenvectors have zero support on the B-sublattice, $u_{0,B, n}=v_{0,B, n}=0$, while the A-sublattice satisfies
\begin{align}
& u_{0,A,n} = r_R^{n-1},\quad
v_{0,A, n}= r_L^{n-1},
\end{align}
where $r_R = -s/(t-\gamma)$ and $r_L = -s/(t+\gamma)$. 
Importntly, the localization ratios $r_{R,L}$ of the TM behave very differently than those of the bulk modes and  in general $r_Rr_L \neq 1$.
In this context, we can identify three different regions for the TM, namely: 
\begin{align}
& \mathrm{(I)}: s+\gamma< t, \quad \lvert r_{R} \rvert < 1, \lvert r_L\rvert < 1, \nonumber \\
 & \mathrm{(II)}:   \sqrt{s^2+\gamma^2}<  t< s+\gamma,\quad 
   \lvert r_R\rvert > 1, \lvert r_L \rvert< \lvert r_R\rvert^{-1}<1, \nonumber \\
 & \mathrm{(III)}:\sqrt{s^2+\gamma^2}> t,\quad \lvert r_R \rvert >1,  1>\lvert r_L\rvert > \lvert r_R\rvert^{-1},\label{regions}
\end{align}
%
%
as highlighted in Fig.~\ref{fig:fig_linear_01}(b).
In region (I) both the right and left eigenvectors of the TM, are located on the same side.
On the other hand, in regions (II) and (III), non-reciprocity dominates, causing these eigenvectors to be located opposite to each other. 
Consequently, we examine their localization: in region (II), the right eigenvector of the TM is less localized compared to its left counterpart, and vice versa in region (III).
From here on, we set $t = 1$ without loss of generality.

\begin{figure}[t!]
\includegraphics[width=\columnwidth]{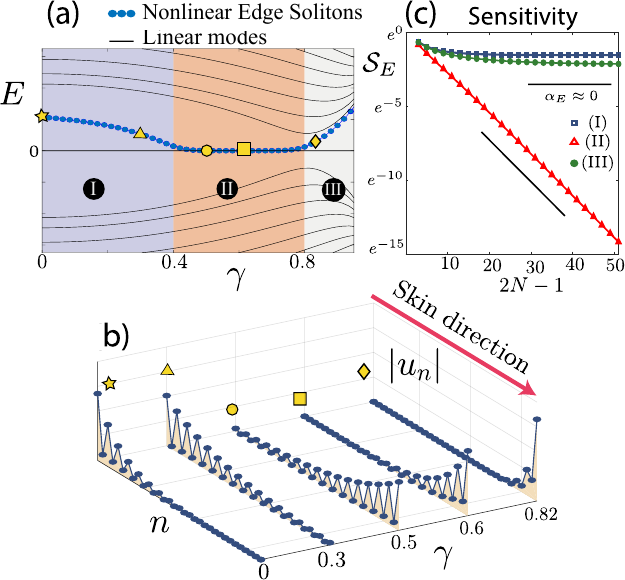}
    \caption{
    (a) Dependence of the energy of the NES at $I=0.5$ against $\gamma$ with $s=0.6$. 
    The black lines are the spectrum of the linear limit.
    (b) Profiles of some representative cases of the NES [see panel (a)].
    (c) Energy sensitivity of the NES as function of the lattice size for (I) $\gamma=0.3$, (II) $0.6$ and (III) $0.825$.
    }    \label{fig:nonlinear_frequency_and_modes}
\end{figure}

We now numerically calculate the nonlinear solutions of Eq.~\eqref{eq:static_equations_of_motion} stemming from the TM, using a pseudo-arclength solver~\cite{DKK1991,DGKMS2008,matcont2023} and follow families of solutions of increasing intensity, i.e.  $I=\sum_j \lvert u_j \rvert^2$ ($j=1, 2, \ldots, 2N-1$) using $\sigma=1$~\cite{ALPHA0002}. 
The solver returns both  the $E$ and the $u_{a,n}$  of the obtained solutions.
Figure~\ref{fig:sensitivity_01}(a) shows the dependence of $E$ as a function of $I$ for representative cases of the families of NES on a chain of $N=17$  with $\gamma = 0.3$ and $0.825$ ($s=0.6$)
in regions (I) and (III) respectively. 
It depicts a behavior similar to the Hermitian SSH model  with $\gamma=0$ [see dashed curve in Fig.~\ref{fig:sensitivity_01}(b)]. 
That is to say, the energy departs from the origin, growing with increasing intensity till approaching the upper band of the linear spectrum when $\sigma>0$, indicating a significant effect of nonlinearity.
This behavior is associated to the SLS/CS breaking induced by Kerr nonlinearity which does not guarantee the presence of solutions at $E = 0$. 

An unexpected result is found in region (II). 
Indeed, despite the aforementioned SLS/CS breaking, we find the family above has $E \approx 0$ when increasing the amplitude, as shown in case of $\gamma=0.6$ and $s=0.6$ in Fig.~\ref{fig:sensitivity_01}(b).
In order to explain the origin of this behavior, 
we rely on the perturbation theory, assuming small amplitude nonlinear solutions originating from the $l$-th eigenmode expand as $\epsilon^{-1/2}\vec{u}_l = \vec{u}_l^{(0)} + \epsilon \vec{u}_l^{(1)} + \mathcal{O}(\epsilon^2)$ with the energy given by $E_l = E_{l}^{(0)} + \epsilon E_{l}^{(1)}+ \mathcal{O}(\epsilon^2)$.
Here the $\epsilon$ is the perturbation parameter which controls the amplitude and the $(\vec{u}_l^{(0)}, E_l^{(0)})$ are retrieved resolving system in its linearized limit.
The details of the first order perturbation analysis $(\vec{u}_l^{(1)}, E_l^{(1)})$ can be found in the supplementary~\cite{SM0001}.
Consequently, we define the rate of change of the energy to the intensity referred to as energy sensitivity factor, 
\begin{equation}
    \mathcal{S}_E^l =  \frac{\partial E_l}{\partial I} \approx\frac{\vec{v}_l^{(0)T}\Gamma_l (\sigma) \vec{u}_l^{(0)}}{\left(\vec{u}_l^{(0)T}\vec{u}_l^{(0)}\right)\left(\vec{v}_l^{(0)T}\vec{u}_l^{(0)}\right)}  = \rho_l E_l^{(1)}(\sigma), 
    \label{peter}
\end{equation}
%
where $\rho_l^{-1}= \vec{u}_l^{(0)T}\vec{u}_l^{(0)}$  and $\{^T\}$ stands for the vector transpose, see supplemental~\cite{SM0001}.
In addition, the $\left[\Gamma_l\right]_{m,j} = \sigma\left\lvert u_{l,m}^{(0)}\right\rvert ^2 \delta_{m,j}$ is the diagonal matrix accounting for the nonlinear perturbation.
According to Eq.(\ref{peter}),
we see that the energy shift stems from two contributions: the nonlinearity through $\delta_l \sim 
 \vec{v}_l^{(0)} \Gamma_l (\sigma) \vec{u}_l^{(0)}$, and the non-orthogonality of the eigenvectors of the underlying linear system quantified by the Petermann-like factor $K_l^{-1}\sim (\vec{u}_l^{(0)T}\vec{u}_l^{(0)})(\vec{u}_l^{(0)T}\vec{v}_l^{(0)})$.

\begin{figure}[t!]
    \centering
    \includegraphics[width=\columnwidth]{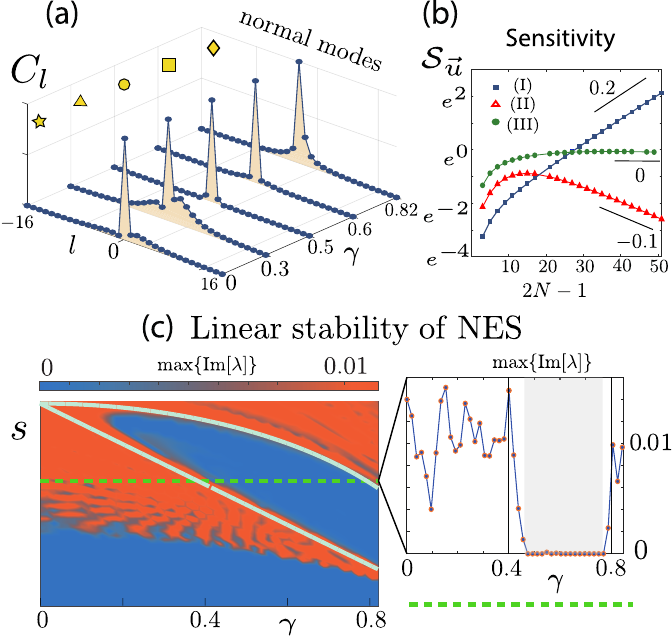}    \caption{
    (a) Projection, $C_l$, of the NES of Fig.~\ref{fig:nonlinear_frequency_and_modes}(b) into the eigenmode basis. The $\lvert C_l\rvert$ are rescaled by their maximum for clarity.
    (b) Similar to Fig.~\ref{fig:nonlinear_frequency_and_modes}(c) but for the wavefunction's sensitivity.
    (c) Linear stability of the NES at fixed $I=0.5$, across the phase diagram of Fig.~\ref{fig:fig_linear_01}(b).
    We display the largest imaginary part of $\lambda$.
    The green solid curves $s = t-\gamma$ and $s^2 = t^2 - \gamma^2$, delimiting the three regions characterizing the TM right and left eigenvectors.
    Right: Dependence of $\max\{\rm{Im}\lambda\}$ against $\gamma$ at $s=0.6$ (see dashed line).
    }
\label{fig:stability_heatmap_01}
\end{figure}

One of the main results of our work is the following relation specific to the NES:
  \begin{equation}
    \mathcal{S}^0_{E} =S_0  
    \frac{\left(r_Lr_R r_Rr_R\right)^N - 1}{\left[ \left(r_R r_R\right)^N - 1\right]\left[ \left(r_L r_R\right)^N - 1\right]} \label{eq:insens},
\end{equation}
with $S_0$ being a positive nonzero constant~\cite{SM0001}. 
This relation suggests that the imbalance between the right and left eigenvectors, which is unique to the TM, i.e. $r_Rr_L\neq 1$, can drastically change the nonlinear energy shift [Eq.~\eqref{eq:insens}] under weak nonlinearity. 
Indeed as the $r_R$ and $r_L$ vary within the parameter space of Fig.~\ref{fig:fig_linear_01}(b), this expression implies that the physical mechanism at the origin of the energy shift of the TM results from the interplay between nonlinearity, $\delta_0 \sim (r_Lr_Rr_R r_R)^N - 1$ and non-orthogonality, $K_0^{-1} \sim [(r_R r_R)^N - 1][(r_L r_R)^N - 1]$.
More specifically, using Eq.\eqref{eq:insens} and looking at large values of $N$, we explicitly get  that in regions (I) and (III) the $\mathcal{S}^0_E \sim 1$. Thus  energy is expected to grow with intensity confirming
the results of Fig.~\ref{fig:sensitivity_01}(a). 
On the other hand, in region (II) we obtain  $\mathcal{S}^0_E \sim \left(r_L r_R\right)^N\equiv e^{-\alpha_E N}$
with $0<\alpha_E(r_R, r_L)<1$, leading to an energy sensitivity which exponentially decays with $N$.
As such, for large lattices we obtain zero-energy NES,
explaining the unexpected numerical observation of Fig.~\ref{fig:sensitivity_01}(b). 

Another way to illustrate this result is obtained by scanning the system's parameter space of Fig.~\ref{fig:fig_linear_01}(b), showing the energy of the NES of fixed nonlinearity, $I=0.5$, varying $\gamma$ along the section with $s=0.6$.
Along this line, the $\gamma = 0.4$ delimits regions (I) and (II) and $\gamma = 0.8$, regions (II) and (III).
The result is shown in 
Fig.~\ref{fig:nonlinear_frequency_and_modes}(a) with the blue dots.
Clearly at $\gamma = 0$ in the Hermitian case, an energy shift away from $E=0$ is seen [star]. 
As $\gamma \rightarrow 0.4$ this energy shift remains non-trivial, while tending to decrease with increasing values of $\gamma$ across region (I).
Representative NES of this region are also shown in Fig.~\ref{fig:nonlinear_frequency_and_modes}(b) for $\gamma = 0$ and $0.3$ respectively the star and triangle dots.
These NES stay localized at the same side as their linear counterparts, yet developing support also on the $B$-sublattice. 
In addition, similar results are also seen for region (III) where $\gamma > 0.8$ [Fig.~\ref{fig:nonlinear_frequency_and_modes}(a)], with a representative NES depicted by the diamond dot in Fig.~\ref{fig:nonlinear_frequency_and_modes}(b). 
These observations, once again, are due to the SLS/CS breaking owning to the Kerr nonlinearity.

Bearing the above in mind, entering region (II) with $0.4<\gamma < 0.8$, we find this family of NES at $I=0.5$ has $E\approx 0$, Fig.~\ref{fig:nonlinear_frequency_and_modes}(a).
In addition, we depict representative NES with $\gamma = 0.5$ and $0.6$ respectively the circle and square dots in Fig.~\ref{fig:nonlinear_frequency_and_modes}(b).
Their profiles clearly show localized shapes, with support also on the B-sublatttice.
Figure~\ref{fig:nonlinear_frequency_and_modes}(c) displays the dependence of the sensitivity factor, $\mathcal{S}_E^0$ against the lattice size $N$.
The measures are obtained as in Fig.~\ref{fig:nonlinear_frequency_and_modes}(a) for representative parameter sets (I) $\gamma = 0.3$, (II) $0.6$ and (III) $0.825$ with $s=0.6$.
For regions (I) and (III), a clear saturation of the $\mathcal{S}_E^0$ to values in the interval $[0.1, 1]$ is seen.
On the other hand, for $\gamma = 0.6$ in region (II), we find an exponentially decaying sensitivity factor, $\mathcal{S}_E^0 \sim e^{-0.6N}$.
It is worth emphasizing that we have also checked that this insensitivity of the NES does not hold for the nonlinear modes emerging from the bulk and are robust whilst the addition of disorder~\cite{SM0001}.


Let us now focus on the shape of the NES rather than their energy. 
The perturbation theory can also be used to define the wavefunction sensitivity
%
\begin{equation}
\mathcal{S}_{\vec{u}}^l = \frac{\partial \vec{u}_l}{\partial I} \approx \rho_l V_l^{(1)}(\sigma), \label{esshape}
\end{equation}
with $V_l^{(1)} = \lvert \sum_{m\neq l} c_{l,m} (\sigma) u_m^{(0)}\rvert$ being a mode overlap integral~\cite{SM0001}. 
Consequently, the $\mathcal{S}_{\vec{u}}^0$ quantifies how strong the TM couples to the bulk modes
when increasing the intensity~\cite{SM0001}.
 We project the NES of Fig.~\ref{fig:nonlinear_frequency_and_modes}(b) onto the eigenmode basis, $\vec{u} = \sum_l C_l \vec{u}_l^{(0)}$ and the corresponding coefficients 
$\lvert C_l \rvert$, are shown in Fig.~\ref{fig:stability_heatmap_01}(a) for  fixed $s$, varying the values of $\gamma$ spanning across regions (I), (II) and (III).
It follows that despite these NES being obtained for the same nonlinearity, $I=0.5$, the TM strongly couples to the bulk [star, triangle and diamond dots] in regions (I) and (III), while in region (II), this coupling is smaller [circle and square dots].
Further, the dependence of $\mathcal{S}_{\vec{u}}^0$ as function of $N$ in Fig.~\ref{fig:stability_heatmap_01}(b) shows that this mode coupling grows or remains constant with increasing $N$ in regions (I) and (III) respectively.
Indeed, we expect that as the lattice size grows, the number of eigenmodes, the TM pairs with, increases.
Remarkably, region (II), evades this anticipation and the $\mathcal{S}_{\vec{u}}^0$ tends to exponentially vanish.
It follows that the TM do not practically couples with the bulk modes in large lattices.
In the supplementary~\cite{SM0001}, we show that these results are independent of $I$.
We also demonstrate that, similar to the TM case, the shape of finite amplitude NES can be modified by tuning  $\gamma$.

To complete our study we move to the linear stability of the NES.
The latter is obtained 
monitoring the time evolution of a small deviation from a nonlinear stationary solution in the form $\psi_{a, n}(\tau) = \left(u_{a, n} + \chi w_{a, n}(\tau) \right)e^{iE\tau}$ with $\chi \ll 1$, $w_{a,n}\propto 1$ and  $\tau$ being the temporal variable. 
After linearization, we obtain a linear evolution equation for the perturbation $w_{a,n}$.
Considering $w_{a, n}(\tau) \sim e^{i\lambda \tau}$ leads to a linear eigenvalue problem $\lambda \vec{w} = Z \vec{w}$~\cite{SM0001,PKF2005}. 
Thus, instabilities are signaled by the imaginary part of $\lambda$.
The largest value of the imaginary part of $\lambda$ is plotted as a colormap in Fig.~\ref{fig:stability_heatmap_01}(c), for values of the parameters of the system throughout Fig.~\ref{fig:fig_linear_01}(b) using the same setting as in Fig.~\ref{fig:nonlinear_frequency_and_modes}.
We observe two regions of linearly stable NES [blue islands].
The stability in the lower islands, is likely due to large band gap, as $s\rightarrow 0$ and the NES being far away from the linear energy bands.
On the other hand, despite region (II), being in the neighbourhood of the band closing, the obtained NES are in general linearly stable, right panel of Fig.~\ref{fig:stability_heatmap_01}(c).

\begin{figure}[t!]
    \centering
   \includegraphics[width=\columnwidth]{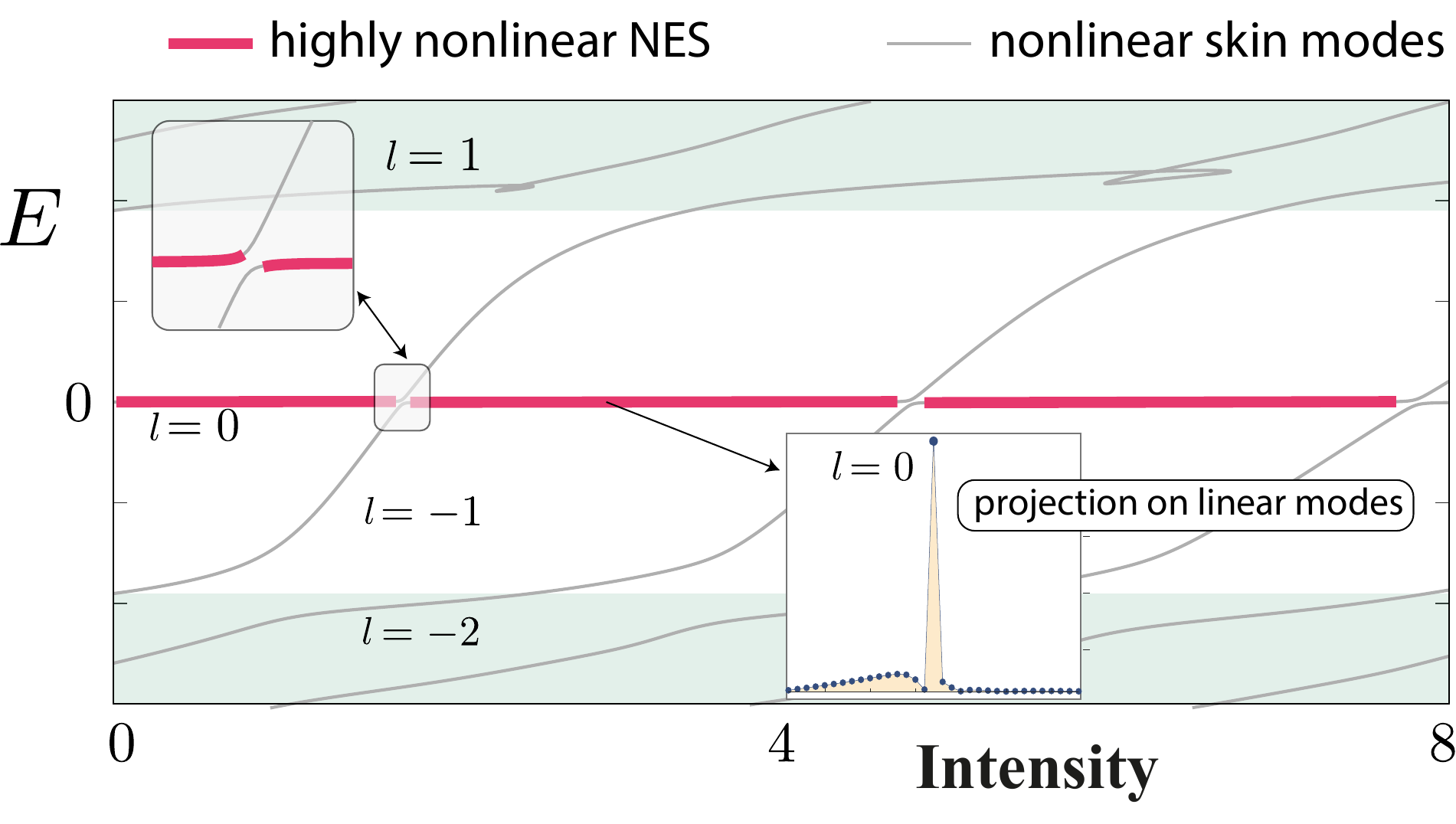}
   \caption{The energy, $E$ against intensity, $I$ for the families of nonlinear modes emerging from all eigenmodes [we show $8$ of them] on a chain of $N=17$ cells with (II) $\gamma = 0.6$ and $s= 0.6$. 
    The insets (left) zoom into a transition and (right) depict the projection of the NES of $E\approx 0$ at $I=2.75$ into the eigenmodes basis (red dots).
    }
\label{fig:nonlinear_spectrum_strong_01}
\end{figure}

To go one step further, we present results  beyond the weakly nonlinear regime and its perturbative analysis. 
To do so we perform numerical continuation like in Fig.~\ref{fig:sensitivity_01} for all eigenmodes with $\gamma=0.6$ and $s=0.6$, in region (II), toward $I\rightarrow \infty$.
The results are shown in Fig.~\ref{fig:nonlinear_spectrum_strong_01}.
It displays  persistent NES with $E\approx 0$ (red lines) even at large intensity. 
In fact, we see that as we increases the nonlinearity from $I=0$, the famlies of NES originating from the TM ($l=0$) at constant $E\approx 0$, and the one emerging from the bulk with $l=-1$ (grey line) of growing $E$, inexorably comes closes around $I=2.5$ without crossing.
We find at this point the characters of the two families are exchanged [left inset of Fig.~\ref{fig:nonlinear_spectrum_strong_01}].
Consequently, the family of NES emerging from the $l=-1$ bulk mode has constants $E\approx 0$ for growing $I$ pass $I=2.5$.
Further, we  
project on the eigenmode basis the high-amplitude NES at $I=2.75$ and find weak couplings of the TM with its surrounding, like what is seen for weak amplitude NES,
right-inset of Fig.~\ref{fig:nonlinear_spectrum_strong_01}.
Furthermore, a cascade of such transitions appears between consecutive families of nonlinear modes emerging from the bulk modes (e.g., $l=-1$ kicks  $-2$, $l=-2$ kicks $-3$, etc).
It would be worthwhile to explore the mechanisms at the origin of these zero-energy highly nonlinear NES, using more precise methods of bifurcation like normal form theory~\cite{K1998}.

In summary, we have investigated the spectral sensitivity of edge solitons arising from the topological modes within a nonreciprocal Su-Schrieffer-Heeger chain featuring local Kerr nonlinearity.
We found a family of nonreciprocal edge solitons (NES) with zero-energy without requiring chiral or sublattice symmetries, as usually the case~\cite{JD2022}, for e.g., when implementing complex nonlinear couplings~\cite{BLLFWX2024,GJPLDF2024}.
 Furthermore, we anticipate that our findings can be extended to other classes of nonlinear localized modes, such as breathers~\cite{A1997,FG2008}.
Consequently, our work may open new ways to manipulate waves in topological systems for applications in sensing, lasing and information processing.

\begin{acknowledgments}
V.A and B.M.M acknowledge the support of the program \emph{Etoiles Montantes en Pays de la Loire} and the  EU H2020 ERC StG ``NASA" Grant Agreement No.
101077954.
The authors would like to express their gratitude to the anonymous reviewers for their comments and suggestions, which helped improve the clarity of this manuscript.
\end{acknowledgments}

\bibliography{bibliography}

\begin{thebibliography}{76}
\bibitem [{NOT()}]{NOTE0001}%
  \BibitemOpen
  \href@noop {} {}\bibinfo {note} {It follows that in the rest of this paper we
  use the biorthogonality relation:
  $\vec{v}_k^T\vec{u}_{k^\prime}=\delta_{kk^\prime}$.}\BibitemShut {Stop}%
\end{thebibliography}

\end{document}